\documentclass[useAMS,usenatbib]{mnras}

\usepackage[T1]{fontenc}
\usepackage{ae,aecompl}
\usepackage{array}

\usepackage{graphicx}
\graphicspath{{figures/}}	
\usepackage{amsmath}	
\usepackage{epstopdf}
\usepackage{multirow}
\usepackage{amssymb}

\usepackage{newtxtext,newtxmath}
\usepackage[normalem]{ulem} 
\usepackage{makecell} 

\newcommand{\msolar}{M$_{\odot}$}

\title[Comparison of star formation history methods]{Comparison of methods used to derive the Galactic star formation history from white dwarf samples}

\author[Roberts et al.]{Emily K. Roberts,$^1$\thanks{E-mail: Emily.Roberts.1@warwick.ac.uk}
Pier-Emmanuel Tremblay,$^1$ Mairi W. O'Brien,$^1$ Antoine Bédard,$^1$ Tim Cunningham,$^2$\thanks{NASA Hubble Fellow} \newauthor{ Conor M. Byrne,$^1$ and Elena Cukanovaite$^1$}
  \\
$^{1}$ Department of Physics, University of Warwick, Coventry CV4 7AL, UK \\
$^{2}$ Center for Astrophysics, Harvard \& Smithsonian, 60 Garden St., Cambridge, MA 02138, USA
}

\date{Accepted XXX. Received YYY; in original form ZZZ}

\pubyear{2024}

\begin{document}
\label{firstpage}
\pagerange{\pageref{firstpage}--\pageref{lastpage}}
\maketitle

\begin{abstract}
We compare three methods of deriving the local Galactic star formation history, using as a benchmark the \textit{Gaia}-defined 40\,pc white dwarf sample, currently the largest volume complete sample of stellar remnants with medium-resolution spectroscopy. 
We create a population synthesis model to 1) reproduce the observed white dwarf luminosity function, 2) reproduce the observed absolute \textit{Gaia G} magnitude distribution, and 3) directly calculate the ages of all individual white dwarfs in the 40\,pc volume. We then compare the star formation histories determined from each method. 
Previous studies using these methods were based on different white dwarf samples and as such were difficult to compare. Uncertainties in each method such as the initial mass function, initial-final mass relation, main sequence lifetimes, stellar metallicity, white dwarf cooling ages and binary evolution are accounted for to estimate the precision and accuracy of each method. We conclude that no method is quantitatively better at determining the star formation history and all three produce star formation histories that agree within uncertainties of current external astrophysical relations. 
\end{abstract}




\section{Introduction}~\label{sec:intro}

The topic of galaxy formation and evolution is one that underpins many different areas of astronomy, from the history of individual galaxies to the specifics of the cosmological model used \citep{Stewart_2008}. Studying the history of the Milky Way is beneficial to understanding similar field galaxies.
From simulations using $\Lambda$\,CDM frameworks, 95 per cent of galaxies like the Milky Way have experienced a small merger event in the last 10\,Gyr \citep{Stewart_2008}, making it likely that the Milky Way will have experienced one too. There has also been evidence of star formation quenching in the Milky Way's history \citep{Haywood_2018} which follows after merger events as shown by simulations using $\Lambda$\,CDM frameworks such as \cite{DiMatteo_2008} or by the presence of a galactic bar \citep{Haywood_2016, Khoperskov_2018}. Merger events can therefore trigger star formation bursts or lulls that could be detectable in the star formation history \citep{Helmi_2018, Antoja_2018}. 

The star formation rate tells us how many stars formed at a given time in a population's history and is one way to investigate the evolution of a galaxy. The star formation rate is a time-varying function that spans the whole lifetime of the population under study and forms the star formation history of the subject; as such, these two terms are often used interchangeably. 
Many different methods have been developed to probe the star formation history on galaxy-wide scales, such as resolving features of colour-magnitude diagrams \citep{Weisz_2008}, measurements of IR dust emission \citep{Rieke_2009}, and near-UV emission measurements to trace recent (in the last 10-200\,Myr) star formation \citep{Salim_2007}. Methods focussed specifically on the Milky Way's star formation history can use more detailed information on individual stars. \cite{Mor_2019}, for example, use simulated colours, magnitudes, and parallaxes compared to \textit{Gaia} Data Release~2 to infer the star formation rate of the Galactic disc. Nucleocosmochronology may also prove a valuable tool in age determinations of individual stars and therefore star formation histories \citep{Ludwig_2010}. There are other methods that use kinematics, isochrones, or asteroseismology (see \citealt{Soderblom_2010} and references therein for a review), providing a plethora of approaches to the field.

Methods that only use white dwarf stars have also been employed, such as observing and modelling the luminosity function \citep{Winget_1987, Rowell_2013}, calculating the ages of individual white dwarfs \citep{Tremblay_2014,Kilic_2019}, and modelling the absolute magnitude distribution \citep{Cukanovaite_2023}. White dwarf stars are the electron degenerate remnants of main sequence stars with masses of less than $8-10$\,\msolar~depending on metallicity \citep{Iben_1997}. Because white dwarfs have no nuclear fuel of their own, they cool predictably with a high dependence on age \citep{Mestel} making them potential chronometers. However, their practicality as chronometers has until recently been hindered by the fact that, by their nature, white dwarfs are faint and harder to observe.

With the advent of \textit{Gaia}, our ability to use white dwarfs in this manner has been revolutionised by the quantity and quality of observational data \citep{Gaia_2016, Gaia_2016b, Gaia_2018, Gaia_2021}. The cataloguing of \textit{Gaia} white dwarfs from Data Release~2 \citep{Barcelona_2018,GentileFusillo_2019} and early Data Release~3 \citep{GentileFusillo_2021} provides far larger samples on which to apply previously established methods as well as opportunities for new methods for estimating stellar formation histories. The full potential of \textit{Gaia} data has yet to be explored.

Despite the advances made possible by \textit{Gaia},
in any full model of a stellar population and its history there remain many systematic uncertainties, both from \textit{Gaia} data itself and other underlying assumptions made in modelling. In particular, even with the same method of determining the star formation history of a population, one may use different astrophysical relations, for example different initial-final mass relations, to map the mass of a main sequence star to its final mass as a white dwarf, or white dwarf evolutionary models. While most previous attempts to derive the star formation history of the solar neighbourhood have included systematic uncertainties in their analyses, significantly different histories were inferred from various white dwarf \citep{SFH2004,Rowell_2013,Tremblay_2014,Fantin_2019,Isern_2019,Cukanovaite_2023} and main-sequence studies \citep{Cignoni_2006,Mor_2019,Alzate_2021,DalTio_2021,Gallart_2024}. It is unclear if these differences arise from different analysis methods and their systematics or different data sets.

In this work, we therefore use the same white dwarf sample to compare three methods of deriving the star formation history of a population: the white dwarf luminosity function, the absolute magnitude distribution and direct age calculations. We use the same external systematic uncertainties from astrophysical relations for each method to investigate whether there is a quantitatively best method, and a consistently best fitting star formation history for the local Galactic population. We present a comparison of the three methods and the best fitting star formation history as chosen by each of the three methods. We also look at the systematic errors that affect modelling white dwarf populations and explore which areas may require future work to help constrain Galactic models further. 

One way to lessen data biases is to use a volume complete sample of stars. 
In this work, we focus on the population within 40\,pc of the Sun, and the local star formation history inferred from white dwarfs \citep{Cukanovaite_2023}. This volume-complete sample of $1073$ white dwarfs from \textit{Gaia} Data Release~3 that have been spectroscopically confirmed was outlined in \cite{OBrien_2024}.

In Section~\ref{sec:40pc} we outline the \textit{Gaia} 40\,pc sample of white dwarfs. Section~\ref{sec:methods} describes the three stellar formation history methods we have compared using the 40\,pc sample. 
Section~\ref{sec:errors} looks at the systematic uncertainties that govern each method, which are discussed in Section~\ref{sec:reduce_err}.
Finally, we discuss the conclusions of this work in Section~\ref{sec:disc}.

\section{The 40 pc sample of white dwarfs}~\label{sec:40pc}

Our sample is comprised of all white dwarf candidates from \textit{Gaia} Data Release~3 in the \cite{GentileFusillo_2021} catalogue, located within 40\,pc of the Sun, and spectroscopically confirmed \citep{OBrien_2024}. This gives a total of 1073 white dwarfs, of which 658 are listed as DA white dwarfs for their primary spectral type (showing hydrogen spectral features), 17 are listed as DB white dwarfs (showing neutral helium spectral features), 285 are featureless DC white dwarfs, while the remaining 113 are a mixture of DQs with detectable carbon features, DZs with detectable metal lines, and DXs which are dependent on individual analysis (see \citealt{OBrien_2024} and references therein for individual spectral types). For all the values listed here, the spectral type listed, e.g. DA, includes all sub-types such as DAZ, DAQ, etc. 
Excluding DC white dwarfs below 5000\,K with unconstrained compositions, 77.5 per cent of the sample have hydrogen dominated atmospheres and 22.5 per cent have helium dominated atmospheres. All white dwarfs have published atmospheric and stellar parameters appropriate for their atmospheric compositions \citep{GentileFusillo_2021,Caron_2023,OBrien_2024,Vincent_2024}. 

If a population of white dwarfs evolves according to single star evolution, we expect a constant median mass \citep{Tremblay_2016} of $\approx0.6$\,\msolar\ (e.g. \citealt{McCleery_2020}) across all temperatures. However, when fitting the white dwarf photometry and \textit{Gaia} parallax, the median mass drops significantly below $\approx6000$\,K, most likely due to inaccuracies in opacities such as the red wing of Ly\,$\alpha$ \citep{Caron_2023} rather than binary evolution causing genuine lower masses. This low mass issue was corrected for in \cite{OBrien_2024} using a fifth-order polynomial to bring \textit{Gaia} masses in line with the expected constant median mass. To predict the same absolute $G$ magnitude, the corresponding effective temperatures were also corrected using the white dwarf mass-radius relation \citep{Bedard_2020}.

The other issue regarding mass determinations in the 40\,pc sample is the appearance of seemingly low mass white dwarfs (below $\approx0.54$\,\msolar) which have predicted total lifetimes longer than that of the Universe. These white dwarfs are understood to be either statistical outliers, unresolved double degenerate systems erroneously assumed to be a single white dwarf in the \textit{Gaia} fitting procedure, or more rarely within a volume, true low mass white dwarfs that went through binary evolution \citep{Cunningham_2024,OBrien_2024}. By having two objects within \textit{Gaia}'s spatial resolution limit, if a single white dwarf with the luminosity of the two objects is assumed, it makes the radius much larger and, consequently, the mass much lower than is physically possible for a single white dwarf within the age of the Universe. Because the vast majority of double degenerate candidates within 40\,pc are unconfirmed and the minimum mass of a single white dwarf is uncertain \citep{Kalirai_2012, Cummings_2018, ElBadry_2018, Marigo_2020,Hollands_2024,Cunningham_2024}, a sharp cut off at 0.54\,\msolar\ to exclude all white dwarfs below this mass was deemed the best compromise for the purpose of method comparison analysis.

\begin{figure}
	\includegraphics[width=\columnwidth]{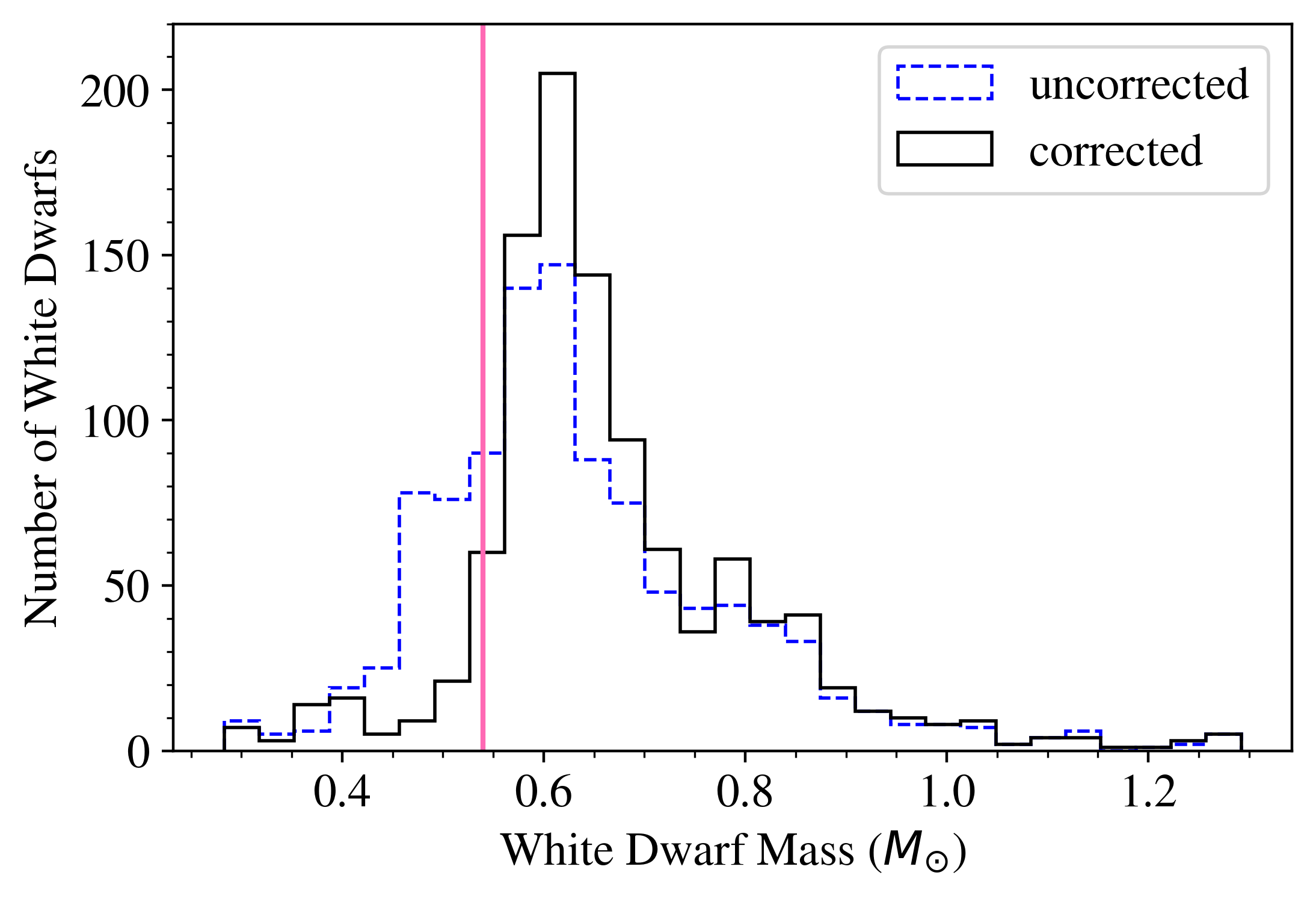}
    \caption{Photometric white dwarf mass distribution of the \textit{Gaia} 40\,pc sample (blue dashed line) compared with the mass distribution after the low mass corrections of \protect\cite{OBrien_2024} for cool white dwarfs ($T_{\rm eff} < 6000$\,K) is applied (black solid line). The vertical (pink) line represents the lower mass cut off, below which the objects are considered to be double degenerate and other binary white dwarfs and are excluded from this work.}
    \label{fig:massdist}
\end{figure}

In Fig.\,\ref{fig:massdist}, we show the mass distribution for the 40\,pc sample with and without the low-mass correction accounting for opacity modelling issues, as well with the vertical double degenerate cut off line at 0.54\,\msolar. The rest of this work was conducted on the sample of 960 objects with corrected masses above 0.54\,\msolar. 

The 40\,pc sample has, as mentioned, both DA and non-DA spectral types within it. There are some advantages to using the DA-only sample, for example the \textit{Gaia} derived masses are more reliable for DA spectral types \citep{Cunningham_2024} since non-DA white dwarfs have trace elements (H, C) that can impact mass determinations \citep{Bergeron_2019, Blouin_2023, Camisassa_2023, Kilic_2025_100pc}. A DA-only sample also reduces the effect of the low mass corrections as DA white dwarfs do not exist below $\approx$5000\,K since H$\alpha$ is not detectable, although this introduces a strong cooling age bias. 
Similarly, a cut-off on lower-than-average mass white dwarfs can reduce uncertainties on stellar parameters by removing objects with long main-sequence lifetimes \citep{Isern_2019,Heintz_2022}, again introducing biases on stellar ages, in part because more massive white dwarfs are more likely to be stellar merger remnants \citep{Temmink_2020}. As our work is attempting to replicate the entire 40\,pc sample and investigate the underlying star formation history, as well as evaluate methods for doing so, we use the full 40\,pc white dwarf sample. 



\section{Methods Under Comparison}~\label{sec:methods}

In order to compare different methods of using white dwarfs to determine the star formation history of a population, we identified three independent methods that could be tested on the 40\,pc sample. 

The first two methods use forward modelling to simulate a synthetic population of white dwarfs with properties that can be quantitatively compared to the observed 40\,pc sample. We look at 1.) the luminosity function (the absolute bolometric magnitude distribution) and 2.) the absolute \textit{Gaia G} magnitude distribution of the sample. Despite using the same simulation to generate the synthetic white dwarf population, the calculations of luminosity/bolometric magnitude and absolute \textit{G} magnitude rely on different subsets of \textit{Gaia} data: luminosity and bolometric magnitude depend on the derived effective temperature and mass whereas absolute \textit{G} magnitude is a function of apparent \textit{G} magnitude and parallax. 

The third employed method infers the age of each white dwarf in the sample from its effective temperature and mass, directly reconstructing the star formation history. By working in reverse, this method is also independent.

The three methods do share many ingredients and systematic uncertainties, such as the initial mass function and initial-final mass relation. The default values of these shared ingredients are listed in Table \ref{tab:ingredients}, while systematic uncertainties on these ingredients are discussed in Section \ref{sec:errors}.

\renewcommand{\arraystretch}{1.5}

\begin{table*}
	\centering
	\caption{The list of default astrophysical relations used to construct the synthetic population of white dwarfs used in the Luminosity Function and Absolute \textit{G} Magnitude Distribution methods, and to calculate the ages of individual white dwarfs in the Direct Age method.}
	\label{tab:ingredients}
	\begin{tabular}{||lll||} 
 \hline
 Ingredient & Key information & Source  \\ \hline \hline
  Population age & 10.6\,Gyr & \cite{Cukanovaite_2023} \\ \hline
  Initial mass range & 0.95--6.84\,\msolar &  \cite{Cunningham_2024}  \\ \hline
  Initial mass function & $\rho(M) \propto M^{-2.35}$ & \cite{Salpeter_1955}   \\ \hline
  Initial metallicity & $Z = Z_{\odot} = 0.0134$ for all stars generated & \cite{Asplund_2009}   \\ \hline
  Main sequence + giant phases lifetimes & Function of mass and metallicity & \cite{Byrne_2024}   \\ \hline
    Initial-final mass relation & Four piece segmented linear fit & This work (see \citealt{Cunningham_2024})   \\ \hline
  He-atmosphere WD fraction & $\approx$0.25 (temperature-dependent) & \cite{OBrien_2024}   \\ \hline
  Merger delays & Probability of forming via merger is a function of mass & \cite{Temmink_2020}   \\ \hline
 Age vs. kinematic relation & $h (\rm scale~height~in~pc) = 10.71(t_\mathrm{total}) + 65$ & \cite{Cukanovaite_2023}   \\
                      & $p(\mathrm{probability~left~volume}) = 1 - 65/h$  &                           \\ \hline
  Cooling models & Theoretical cooling sequences of C/O core white dwarfs & \cite{Bedard_2020}   \\ \hline
  Crystallisation cooling delay & Extra 0.5\,Gyr to cooling at crystallised mass fraction of 0.5 & \cite{Blouin_2020,Kilic_2020}   \\ \hline

 \end{tabular}
\end{table*}

\subsection{Luminosity function}~\label{sec:LF}



In our simulations, 30\,000 stars are generated according to the initial mass function \citep{Salpeter_1955}. Their main sequence and post-main sequence lifetimes up to and including the asymptotic giant branch phase are then assigned according to the latest single-star Binary Population and Spectral Synthesis \citep[BPASS,][]{Eldridge2017,Stanway2018} detailed stellar evolution models presented in \citet{Byrne_2024}, assuming solar metallicity for all stars. 
To assign each star a formation time, a star formation history has to be assumed. Formation times are sampled from an assumed star formation history from a lookback time of $t=0$ (the present day) to $t=t_\mathrm{pop\,age}$ (the population age of the simulation, 10.6\,Gyr by default, see \citealt{Cukanovaite_2023}) to give each star an age.

White dwarfs are assigned masses based on a newly derived, self-consistent initial-final mass relation where our methodology closely follows the one outlined in \citet{Cunningham_2024}. The initial-final mass relation is constructed using the derived \textit{Gaia} masses for the entire 40\,pc white dwarf sample. However, since \citet{Cunningham_2024} used main-sequence lifetimes from \citet{Hurley_2000}, here we re-calculated the initial-final mass relation using the same code but swapping for the BPASS stellar lifetimes. This ensures self-consistency, e.g. for our maximum population age (10.6\,Gyr) and minimum white dwarf mass for single star evolution (0.54\,M$_{\odot}$), the BPASS lifetime of the corresponding main-sequence star cannot be larger than 10.6\,Gyr. We note that this is a significant difference compared to the stellar population synthesis work of \citet{Cukanovaite_2023} for the 40\,pc sample, who used the initial-final mass relation of \citet{ElBadry_2018}.

\begin{figure}
	\includegraphics[width=0.9\columnwidth]{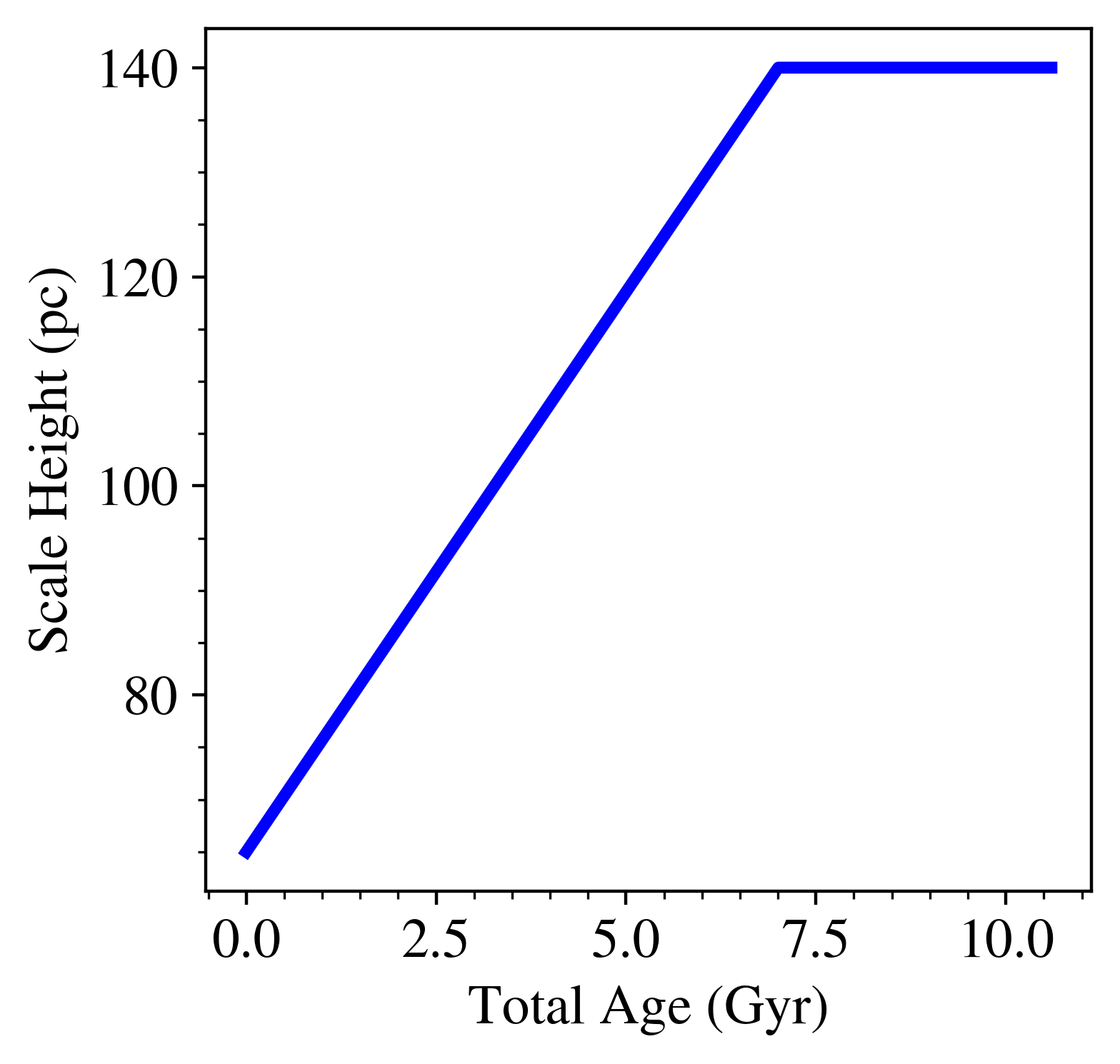}
    \caption{Scale height as a function of the total age of a star. This linear relationship is derived in \protect\cite{Cukanovaite_2023} and flattens off at 140\,pc due to a lack of evidence for kinematic evolution of stars older than around 7\,Gyr \citep{Seabroke_2007}.}
    \label{fig:scaleheight_age}
\end{figure}

It is known that older groups of stars have a higher velocity dispersion in all three Galactic coordinates \citep{Rowell_2019} and so are more likely than younger stars to have left the local 40\,pc volume, which is within the Galactic plane where open clusters dominate stellar formation. \cite{McCleery_2020} only found four halo white dwarfs in the 40\,pc sample, for example. We assume that equal numbers of stars enter and leave the volume in the simulated $x$ and $y$ directions parallel to the Galactic plane. Only in the $z$ direction, perpendicular to the Galactic disc, stars will be leaving and entering the volume at different rates. This allows us to use a one dimensional correction. Using the calculations detailed in \cite{Cukanovaite_2023} to relate the total age of a star, $t_\mathrm{total}$ in Gyr, to its scale height, $h$ in pc, we calculate the probability $p$ of a star having left the volume as:
\begin{equation}\label{eqn:scale_height}
    h = 10.71(t_\mathrm{total}) + 65~,
\end{equation}
\begin{equation}\label{eqn:p_left}
    p(\mathrm{left}) = 1 - 65/h~.
\end{equation}
The derived scale height as a function of age is shown in Fig.\,\ref{fig:scaleheight_age}. The flattening of this function at 140\,pc is due to a lack of evidence for kinematics evolution of stars older than around 7\,Gyr \citep{Seabroke_2007}. 
Only stars that have not left the 40\,pc volume over the course of their lifetime are kept in the simulation.

\begin{figure}
	\includegraphics[width=\columnwidth]{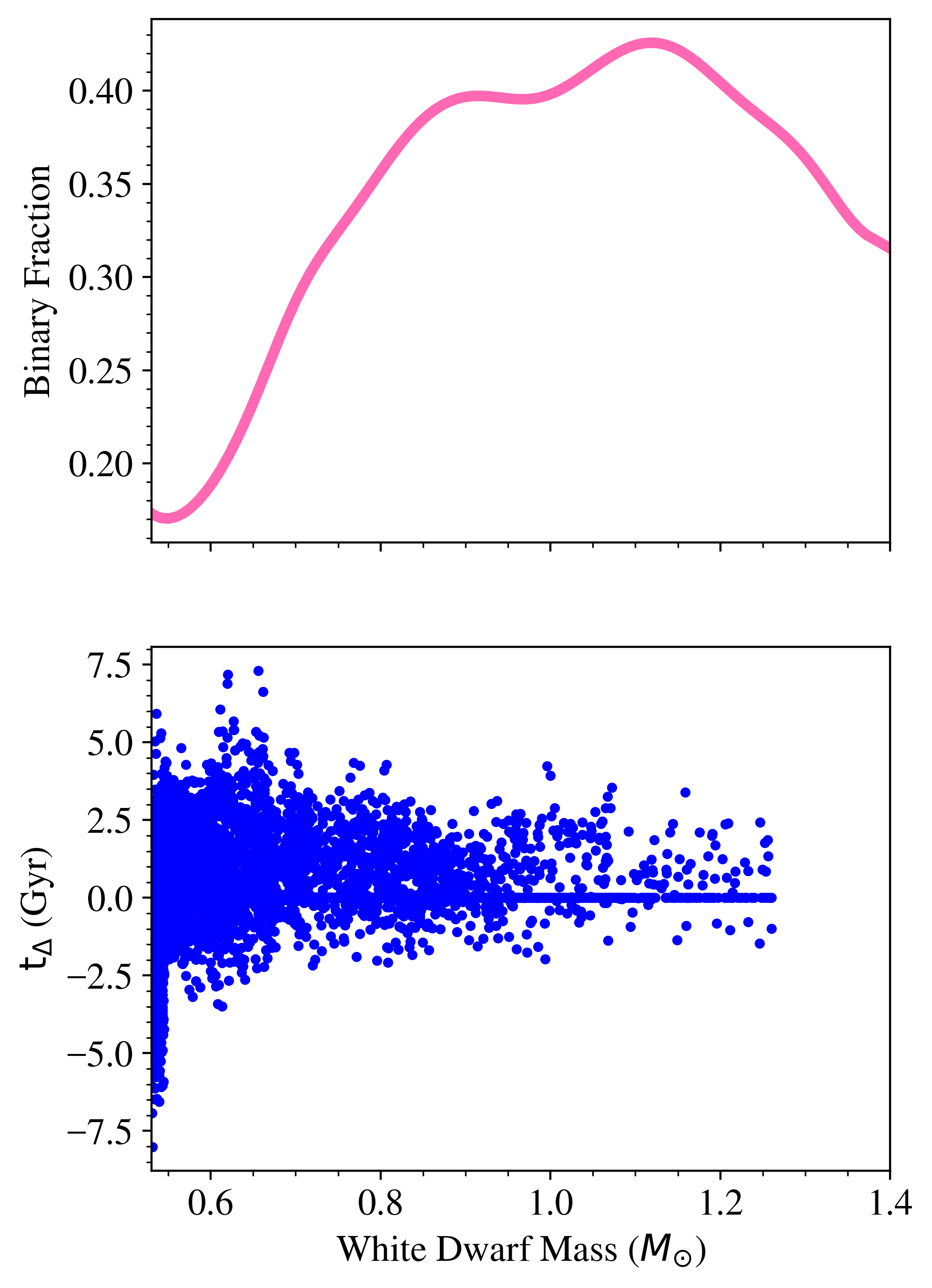}
    \caption{\textit{Top:} The fraction of white dwarfs that formed with a merger in their past as a function of white dwarf mass. This is the default model of \protect\cite{Temmink_2020}. \textit{Bottom:} The time differences in Gyr between the evolutionary history of a white dwarf assuming single star evolution and binary evolution for the stars in the default synthetic population of this work assuming constant formation history. If a star was not formed via binary evolution, the time difference is zero. If it did form via binary evolution and the time difference is positive, the cooling time of the white dwarf is reduced. If it is negative, the cooling time is increased. The majority of the time, $t_\mathrm{\Delta}\geq0$ and a delay is applied.}
    \label{fig:binary_info}
\end{figure}

White dwarfs that are remnants of binary mergers are predicted to make up $10-30$ per cent of single observable white dwarfs \citep{Temmink_2020} and only more recent simulations of white dwarf populations have included such an effect \citep{Toonen_2017,Kilic_2020,Cukanovaite_2023}. This variable is included in our work through the default models of binary populations of \cite{Temmink_2020}, primarily the merger fraction as a function of white dwarf mass, and the differences in inferred white dwarf formation time as a function of white dwarf mass. For each simulated white dwarf, we calculate the probability of it having formed via binary merger of any type, and then, if it did form this way, we sample the delay this would apply to its cooling from a Gaussian distribution with the mean and standard deviation taken from the quartiles shown in Fig. 6 of \cite{Temmink_2020}. The probability of a star forming via binary merger is dependent on mass, as shown in the top plot of Fig.\,\ref{fig:binary_info}, and the cooling delay applied to each star in the simulation can be seen in the bottom plot of Fig.\,\ref{fig:binary_info}. The blue points in the bottom plot show that this cooling delay is not always a delay: the majority of the time the difference is positive and it decreases the white dwarf cooling time for the same assumed formation history; less commonly it is negative and increases the cooling time. Accounting for binary evolution produces on average hotter and brighter white dwarfs for the same formation history than when compared to a population with binary evolution unaccounted for.

To determine which stars in the simulation have become white dwarfs by the present day, the white dwarf cooling time for each star is calculated as:
\begin{equation}~\label{eqn:cooling_time}
    t_\mathrm{WD} = t_\mathrm{total} - t_\mathrm{MS+GP} - t_\Delta
\end{equation}

{\noindent}where $t_\mathrm{WD}$ is the cooling time of the white dwarf, $t_\mathrm{total}$ is the formation time of the star and its total age, $t_\mathrm{MS+GP}$ is the main sequence and giant phases lifetime of the star, and $t_\Delta$ is the change in cooling time due to a merger history. Stars with a negative white dwarf cooling time have not had enough time since their formation to evolve into white dwarfs by the present day, and are discarded from the simulation. 

It is worth noting that we also deal with unresolved binarity in two additional ways: as previously mentioned, we employ a mass cut off of 0.54\,\msolar~that removes the majority of unresolved double degenerate systems from the observed 40\,pc sample; also, unresolved white dwarf and main sequence pairs are selected against in the observational sample as it is constructed from the \textit{Gaia} white dwarf catalogue of \cite{GentileFusillo_2021} which does not select white dwarfs with a large red excess from a main-sequence companion \citep{OBrien_2024}.

The cooling models of \cite{Bedard_2020} depend on the envelope composition of the white dwarf, either hydrogen-rich or hydrogen-deficient. 
The cooling models assume a carbon/oxygen core (50/50 proportions in mass), a helium mantle ($M_{\mathrm{He}}/M_{\mathrm{WD}} = 10^{-2}$), and either a thick or thin hydrogen outer layer ($M_{\mathrm{H}}/M_{\mathrm{WD}} = 10^{-4}$ or $10^{-10}$, respectively).
White dwarfs with standard thick hydrogen layers retain a hydrogen-rich atmosphere for their entire evolution, while hydrogen-deficient white dwarfs exhibit a variable atmospheric composition (and thus spectral type) with cooling age \citep{Bedard_2024}. In this work, we follow the usual convention of modelling hydrogen-atmosphere objects with thick-layer cooling models, and helium-atmosphere objects with thin-layer cooling models. While this does ignore some elements of spectral evolution, given that some thin-layer white dwarfs have hydrogen-rich atmospheres, the simplifying assumption used here has a negligible impact on the 40\,pc sample simulations as fewer than 1\% of the sample are thin-layer white dwarfs with hydrogen-rich atmospheres.

The evolution of the fraction of helium-atmosphere white dwarfs with temperature is determined by Fig.\,6 of \citet{OBrien_2024}. The increasing number of helium-atmospheres with decreasing $T_{\rm eff}$ is thought to originate from the convective mixing of a thin superficial hydrogen layer in the larger underlying helium layer \citep{Rolland_2018, Cunningham_2020, Bedard_2022_STELUM}. Regarding spectral types and predicted colours, we implement this atmosphere composition versus $T_{\rm eff}$ dependence in a probabilistic way, giving each white dwarf the appropriate chance of having a helium or hydrogen-dominated atmosphere.


The cooling models define the crystallised fraction of the white dwarf core. Rather than cooling continuously with time, a crystallising white dwarf will plateau in effective temperature and luminosity as energy released by crystallisation and related processes (latent heat, carbon/oxygen phase separation, and $^{22}$Ne distillation) halts the cooling \citep{Tremblay_2019,Blouin_2021,Saumon_2022,Bauer_2023}. While distillation has been found to be the main cause of the delay in the cooling process, the exact distillation delay for white dwarfs less massive than $\approx$ 1\,\msolar\ is not well understood, except that the delay seems to be peaked at a crystallised fraction of $\approx$ 0.5 \citep{Kilic_2020,Blouin_2020,Bedard_2024_Qbranch}.
In our simulation, for white dwarfs that have a crystallised fraction of $\geq0.5$ at the present day, a distillation delay of 0.5\,Gyr is added to the cooling time since this process is not accounted for in the models of \citet{Bedard_2020}. The choice of this value is justified in Sect.~\ref{sect:high_mass}. 
As the variation of this delay within the white dwarf population is largely unconstrained, we choose to add a small delay to all white dwarfs as an average; we do not rule out the possibility of a larger delay applying to a smaller proportion of the white dwarf population.

\begin{figure}
	\includegraphics[width=\columnwidth]{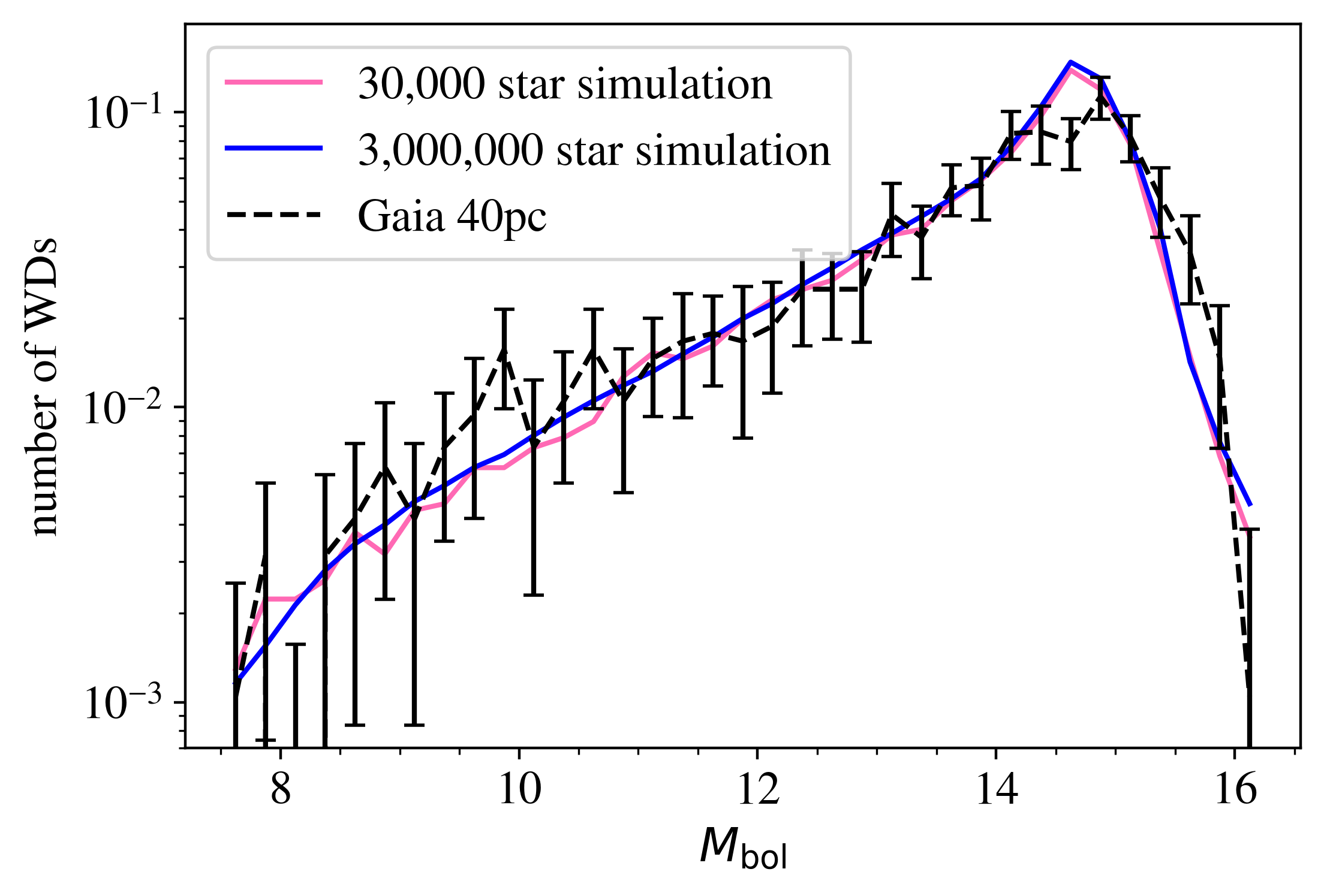}
    \caption{Observed luminosity function (black dashed line) for the 40\,pc white dwarf sample compared to that of the synthetic population of white dwarfs used as the default simulation (pink solid line). A simulation including 3\,000\,000 stars is shown in the solid blue line. The errors on the observed 40\,pc luminosity function are Poisson errors and propagation of errors from \textit{Gaia} uncertainties on mass and effective temperature. The default synthetic luminosity function is generated with a finite number of stars therefore small oscillations in the rising slope are artifacts of the simulation process and not necessarily real features, however the 30\,000 star simulation has converged (as seen by the similarity to the 3\,000\,000 star simulation) and so using 30\,000 stars is suitable.}
    \label{fig:LF}
\end{figure}

Interpolation of these cooling model grids across mass, cooling time, and corresponding envelope composition gives the luminosity of each white dwarf remaining in the synthetic population. To plot the luminosity function of the synthetic population, we convert the luminosities into bolometric magnitudes, $M_{\rm bol}$ and bin the stars by bolometric magnitude. The simulated luminosity function using a constant star formation history in Fig.\,\ref{fig:LF} shows the characteristic features of the white dwarf luminosity function: the rising slope towards fainter magnitudes, a peak at $M_\mathrm{{bol}}\approx 15$, and then a sharp drop off due to the finite age of the population. The luminosity function is scaled to have the same number of total objects as the observed 40\,pc sample. 

The model-dependent observed luminosity function is calculated based on the effective temperatures and masses determined by \citet{OBrien_2024} for the 40\,pc white dwarf sample with the low mass correction applied. These parameters are based on fits of \textit{Gaia} $G$, $G_{\rm BP}$, and $G_{\rm RP}$ magnitudes as well as the parallax with the appropriate model atmospheres. $M_{\rm bol}$ is then calculated using the cooling models of \citet{Bedard_2020}. The \textit{Gaia} luminosity function has error bars that include both Poisson errors from the number of stars in each bin ($\sqrt{N}$ where $N$ is the number of objects in that bin) and propagated errors due to \textit{Gaia} uncertainties on mass and effective temperature.

The reduced $\chi^2$ ($\chi_\nu^2$) for the simulated luminosity function is calculated against the observed function from the 40\,pc sample using:
\begin{equation}
    \chi^2 = \sum_i^N \frac{(O_i - E_i)^2}{\sigma_i^2}
\end{equation}
\begin{equation}
    \chi^2_\nu = \frac{\chi^2}{\nu}
\end{equation}
{\noindent}where the number of degrees of freedom $\nu = n - m$ is the number of observations $n$ (here, the number of bins used in the histogram) minus the number of fitted parameters $m$ (one, for the shape of the histogram overall), $O_i$ is the observed frequency in the $i$th bin, $E_i$ is the expected frequency in the $i$th bin, and $\sigma_i$ is the error on the $i$th observation.



\subsection{Absolute \textit{Gaia G} magnitude distribution}~\label{sec:absG}

The second method explored in this work is to compare the absolute \textit{Gaia G} magnitude distribution to that of the synthetic population. This uses the same population synthesis method as described in the previous section until the point at which the bolometric magnitudes of the individual white dwarfs are calculated. It then diverges and uses the luminosity and effective temperature, both derived from the cooling grids of \cite{Bedard_2020}, combined with model atmosphere grids (\citealt{Tremblay_2011} for the pure hydrogen models and \citealt{Cukanovaite_2021} for the helium and mixed atmosphere models\footnote{We employ the same rules as in \citet{OBrien_2023,OBrien_2024} for the use of mixed versus pure-helium model atmospheres.} where H/He = $10^{-5}$) to calculate the absolute \textit{Gaia G} magnitude for each star. 

Compared to the luminosity function method, the model atmospheres are now used in the population synthesis instead of in the derivation of individual $T_{\rm eff}$ and mass from the observations. Most importantly, only observed \textit{Gaia G} and parallax are used in the absolute \textit{Gaia G} magnitude method, whereas \textit{Gaia} $G$, $G_{\rm BP}$, and $G_{\rm RP}$ magnitudes as well as the parallax are used in the luminosity function method, when data are fitted to obtain $T_{\rm eff}$ and mass.

\begin{figure}
	\includegraphics[width=\columnwidth]{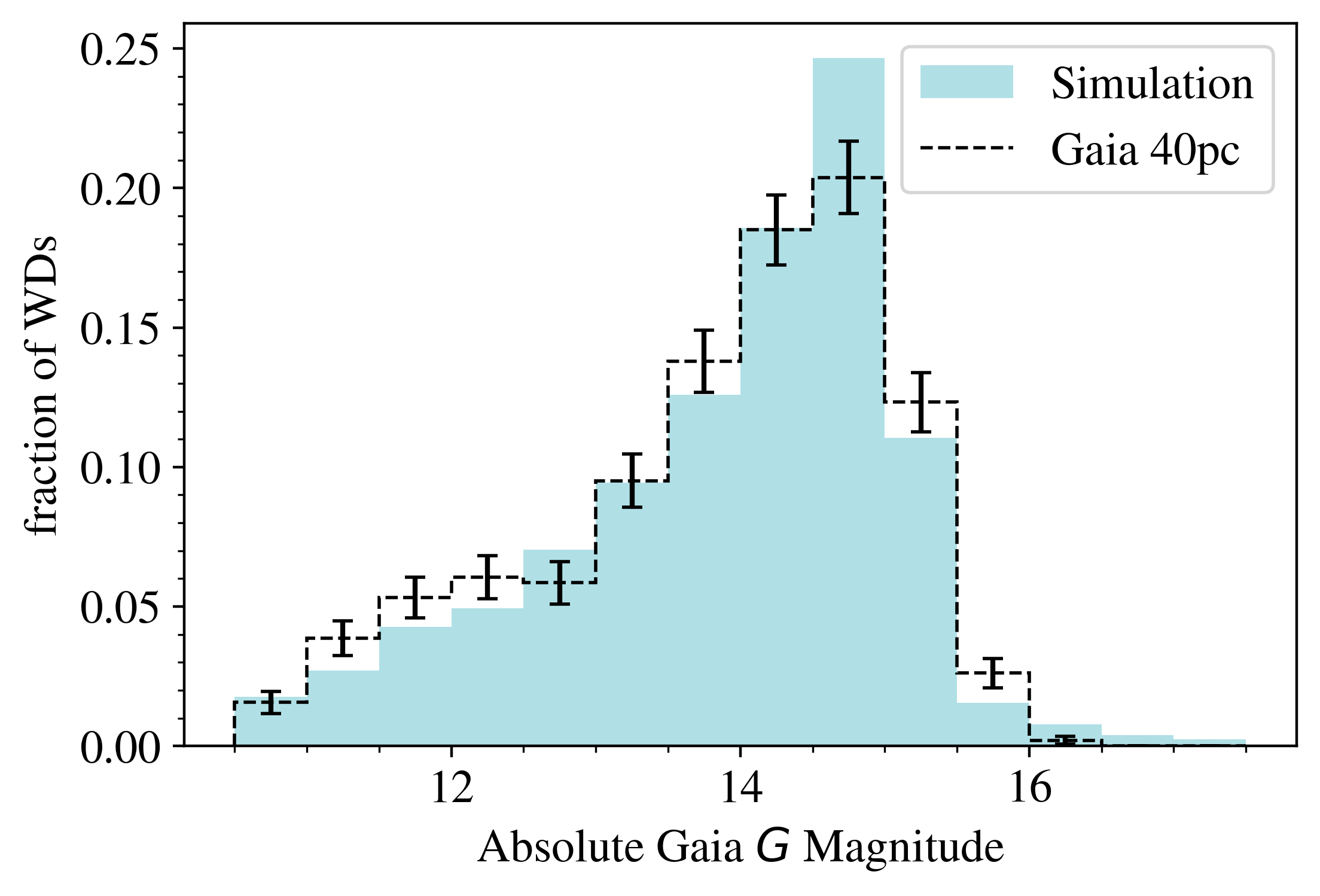}
    \caption{Observed absolute \textit{G} magnitude distribution (black dashed line) for the 40\,pc sample compared to that of the synthetic population of white dwarfs (light blue histogram). The errors on the observed 40\,pc absolute \textit{G} magnitude distribution are Poisson errors.}
    \label{fig:absG}
\end{figure}

The simulated absolute \textit{G} magnitude distribution assuming a constant star formation history in Fig.\,\ref{fig:absG} shows a peak at $\mathrm{M_G}\approx15$\,mag, a drop off at fainter magnitudes and evidence of a bright magnitude shoulder at $\mathrm{M_G}\approx12.5$\,mag. We calculate the $\chi_\nu^2$ to quantify the difference with the observed distribution directly drawn from \textit{Gaia} Data Release~3.


\subsection{Direct age calculations}~\label{sec:DA}

The direct calculation of the age of each white dwarf in the 40\,pc sample requires no comparison to a simulated white dwarf population and uses the \textit{Gaia} derived corrected effective temperatures, masses and compositions as defined by \cite{OBrien_2024}. These white dwarf parameters are then converted to ages using the same models and astrophysical relations as in the previous methods. This method is based on the work of \cite{Tremblay_2014}, although it is adapted for the 40\,pc sample.

The cooling models of \cite{Bedard_2020} are employed to calculate the white dwarf cooling time. If the white dwarf has a crystallised mass fraction $\geq$0.5, it is assumed to have experienced a distillation delay to its cooling and the white dwarf cooling age is increased by 0.5\,Gyr. The pre-white dwarf mass is calculated using our modified initial-final mass relation based on the method of \cite{Cunningham_2024} and then the combined main sequence and post-main sequence lifetime is calculated under an assumption of solar metallicity with BPASS models \citep{Byrne_2024}. Since there is no direct evidence for individual white dwarfs in the 40\,pc sample being the product of mergers or not, as the large majority of them have disc kinematics appropriate for their estimated ages \citep{McCleery_2020,Cukanovaite_2023}, a probabilistic approach to the merger time delay is taken. Each white dwarf is assigned a probability of being formed via merger, and those that are selected as mergers sample the cooling delay range accordingly \citep{Temmink_2020}. The total age of the white dwarf is then calculated as:
\begin{equation}
    t_\mathrm{total} = t_\mathrm{WD} + t_\mathrm{MS+GP} +t_\Delta
\end{equation}
{\noindent}If the inclusion of a merger delay increased the total age of the white dwarf to greater than 10.6\,Gyrs, the merger delay was ignored in the calculation of the total age.

\begin{figure}
	\includegraphics[width=\columnwidth]{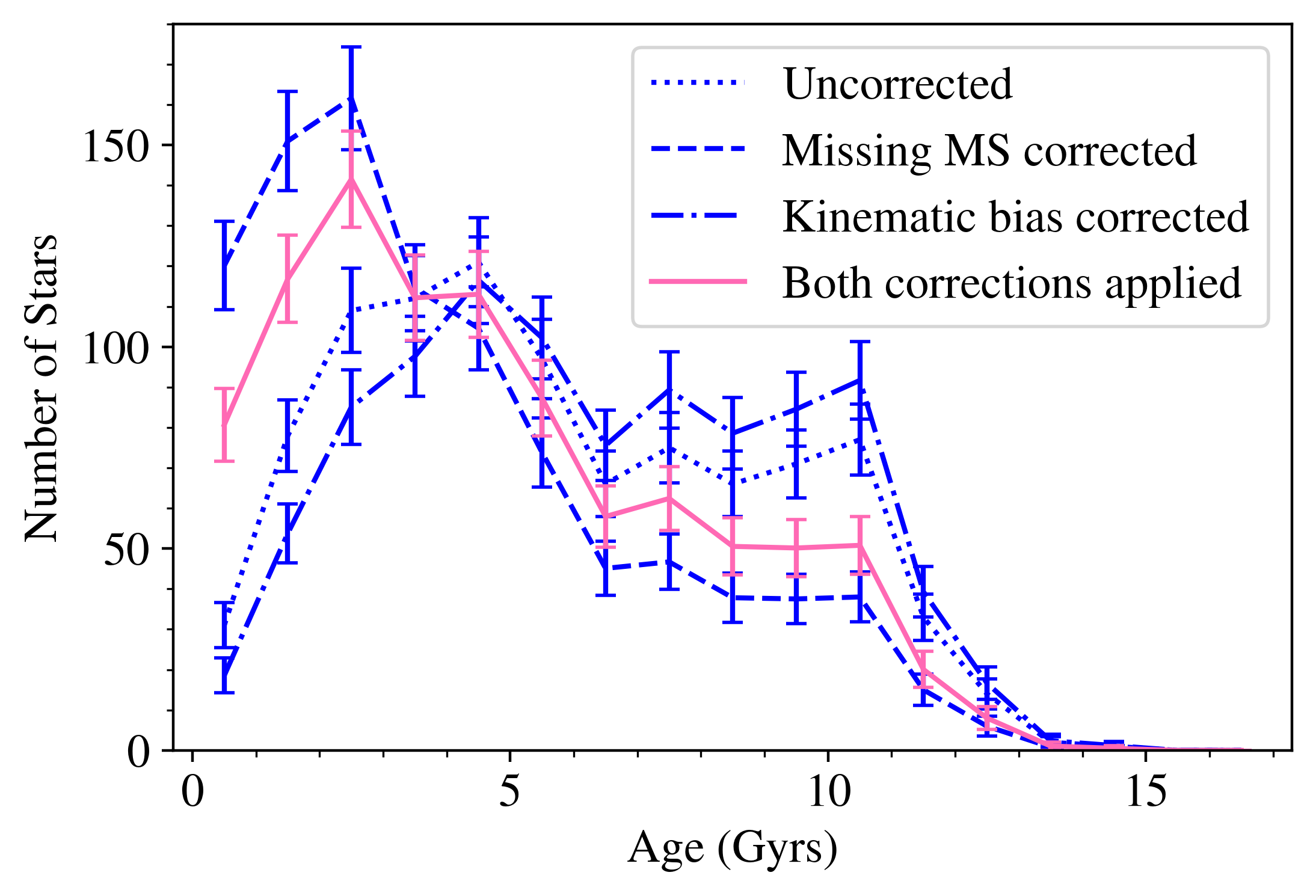}
    \caption{Number of white dwarfs in each 1\,Gyr age bin with the uncorrected values in the blue dotted line and the values with both the main sequence and kinematic corrections applied in the solid pink line. Uncertainties on each bin are Poisson errors.}
    \label{fig:direct_age_result}
\end{figure}

The ages of all the white dwarfs in the 40\,pc sample are then binned to construct an age distribution for the population. Fig.\,\ref{fig:direct_age_result} shows this histogram in a dotted blue line. The uncertainties on these bins are Poisson errors. However, this distribution is not directly the star formation history of the underlying stellar population, as it does not account for stars that will have formed in the population but that are still on the main-sequence or giant phases. It also does not account for biases due to older stars that formed in the 40\,pc sample being more likely to have left the volume due to their higher space velocities on average.

To correct for the missing main sequence stars, the ratio of stars that remain main sequence stars to stars that have become white dwarfs by the present day is calculated from the Salpeter initial mass function according to Eq.\,1 from \cite{Tremblay_2014}:
\begin{equation}
    B_\mathrm{MS-WD} = \frac{\int^{M_\mathrm{lim}}_{M_0} M^{-2.35} dM}{\int^{M_{1}}_{M_\mathrm{lim}} M^{-2.35} dM}
\end{equation}


{\noindent}where $M_0$ and $M_1$ are any mass limits outside the range being studied (0.6 and 10\,\msolar~respectively -- as we are only interested in relative star formation history) and $M_\mathrm{lim}$ is the mass at which $t_\mathrm{lookback} = t_\mathrm{MS+GP}$, that is at a certain lookback time, stars of that mass have just left the asymptotic giant branch and become white dwarfs. This ratio is computed for the lookback time corresponding to the centre of each histogram bin and the number of stars in each histogram bin is adjusted to add back in these missing main sequence stars. The total number of stars is then renormalised to the total in the 40\,pc sample. This corrected age distribution can be seen in Fig.\,\ref{fig:direct_age_result} with the dashed blue line.

To correct for the kinematic bias meaning that older stars are more likely to have left the 40\,pc volume, the probability of a star remaining in the sample is calculated using Eqns.\,(\ref{eqn:scale_height})-(\ref{eqn:p_left}). For each bin, the number of stars is divided by the probability of remaining to get the original number of stars in the bin at that lookback time. As with the main sequence correction, the total number of stars is then renormalised to the total in the 40\,pc sample. This kinematic-corrected distribution can be seen in the dot dashed blue line in Fig.\,\ref{fig:direct_age_result}.

Figure\,\ref{fig:direct_age_result} shows the derived star formation history from this method with both corrections applied to the 1\,Gyr width bins in a solid pink line. The resulting star formation history form finds the first stars to have formed more than 16\,Gyr ago -- even after having removed white dwarfs with masses less than 0.54\,\msolar. These impossibly old stars could be outliers because of  \textit{Gaia} statistical uncertainties or remaining model atmosphere systematic uncertainties even after the ad-hoc mass and $T_{\rm eff}$ correction for $T_{\rm eff} < 6000$\,K.
Due to lower mass white dwarfs having large lifetime uncertainties \citep{Heintz_2022}, statistical age uncertainties on the oldest bins can be up to several Gyr.
These impossibly old stars have an age within 3\,$\sigma$ of 10.6\,Gyr from \textit{Gaia} uncertainties alone, even before we factor in systematic uncertainties from other sources, making them compatible with the 10.6\,Gyr age of the Galactic disc assumed throughout.
We also note that the impossibly old stars are all white dwarfs with masses very close to the 0.54\,$M_{\odot}$ limit.
We observe a qualitative onset of stellar formation at lookback times of 10-12\,Gyr, which is consistent with the age of the disc of 10.6\,Gyr estimated by \citet{Cukanovaite_2023} from the same sample. The star formation rate appears to increase from the onset of star formation, reaching a peak at 2--3\,Gyr ago. The star formation rate appears to be about 2.5 times higher at recent times (less than 6\,Gyr ago) compared to at older times.

\subsection{Testing of star formation history forms}~\label{sec:sfhs}

\begin{figure}
	\includegraphics[width=\columnwidth]{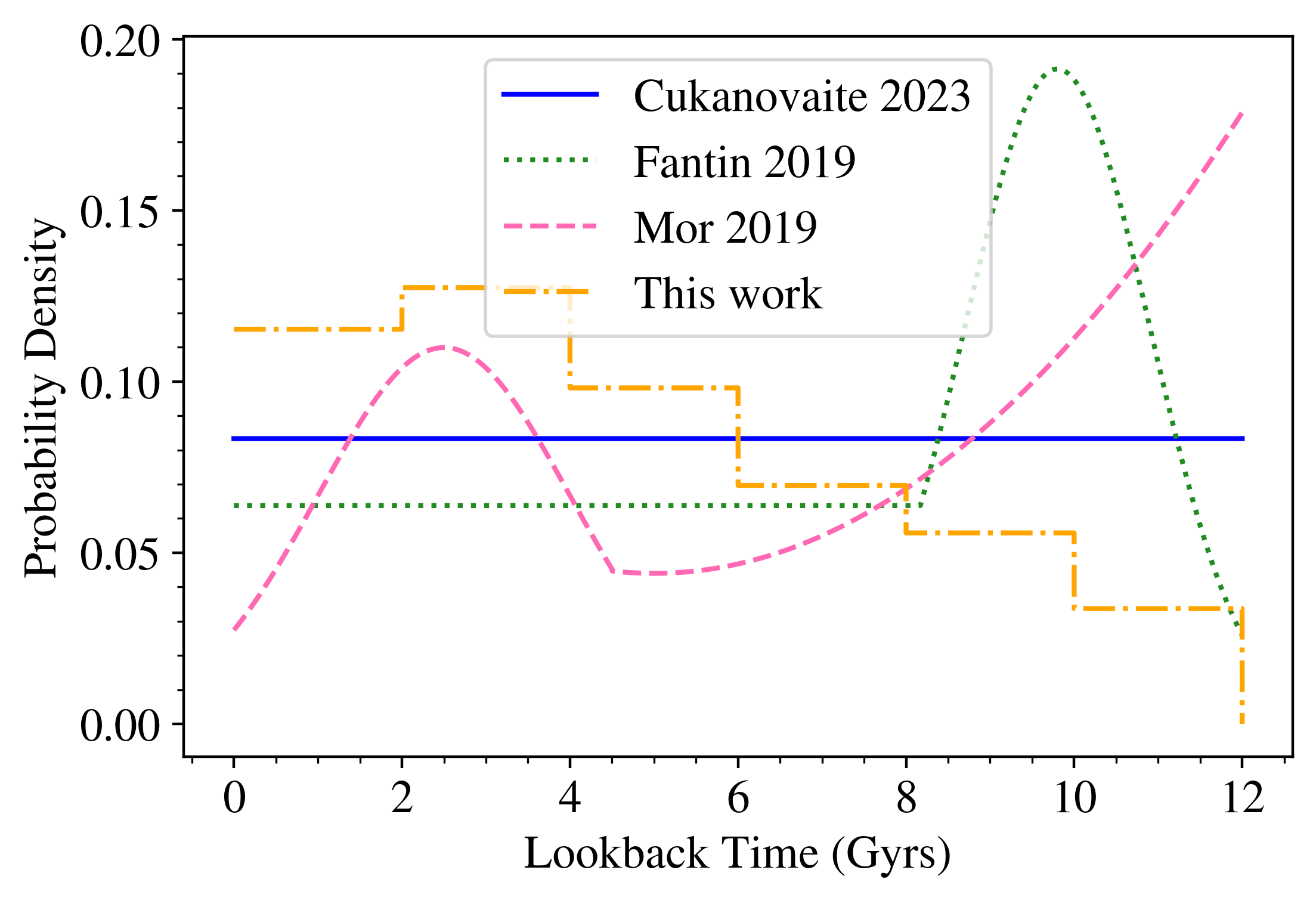}
    \caption{The four forms of star formation history used throughout this work. The four forms are approximated from \protect\cite{Cukanovaite_2023}, \protect\cite{Fantin_2019}, \protect\cite{Mor_2019} and the results of our direct age method. Each form has a very different shape or peaks and allows a wide range of options for our tests.}
    \label{fig:sfh_forms}
\end{figure}

Across many data sets, methods and stellar populations, there are many derivations of the local star formation history. Two of the methods discussed in this work (the luminosity function and the absolute \textit{G} magnitude distribution) rely on an underlying assumption of a star formation rate to generate a synthetic population of white dwarfs. The third method, calculating the ages of individual white dwarfs directly, gives a star formation history by itself. To test whether all methods are consistent, we select four forms of the star formation history from previous publications to test our methods on: 1) \citet{Cukanovaite_2023} which finds a constant star formation rate for the last 10.6\,Gyr, 2) \citet{Mor_2019} that has a star formation rate with two peaks ($\approx2.5$ and 10\,Gyr ago), 3) \cite{Fantin_2019} that has a strong peak 9.8\,Gyr ago and finally 4) we feed back our results from the direct age method (Fig.\,\ref{fig:direct_age_result}), which has a enhanced star formation at recent times ($\lesssim 5$ Gyr). The forms of the four star formation histories as they are approximated in this work are shown in Fig. \ref{fig:sfh_forms}.


We ran luminosity function simulations with each of the four different assumed star formation history forms and calculated the $\chi^2_\nu$ for each with the default choices for each of the ingredients as listed in Table\,\ref{tab:ingredients}. The $\chi^2_\nu$ values of these runs can be seen in Table\,\ref{tab:best_chi}.

\renewcommand{\arraystretch}{1.5}

\begin{table*}
	\centering
	\caption{The $\chi_\nu^2$ value found for each of the four star formation history forms when used in the Luminosity Function and Absolute \textit{G} magnitude distribution methods as compared with the observed 40\,pc sample. These simulations all use the default values for simulation ingredients as described in Table \ref{tab:ingredients}. The four star formation history forms are the approximated versions described in Sect. \ref{sec:sfhs} and shown in Fig.\,\ref{fig:sfh_forms}.}
	\label{tab:best_chi}
	\begin{tabular}{||l|l|l||} 
 \hline
  Star Formation History Form & $\chi^2_\nu$ Luminosity Function & $\chi^2_\nu$ Absolute \textit{G} Magnitude Distribution   \\ \hline \hline
  Cukanovaite et al. 2023 & 3.15 & 2.85 \\ \hline
  Mor et al. 2019 & 2.47 & 2.99 \\ \hline
  Fantin et al. 2019 & 4.19 & 9.43 \\ \hline
  This work (Direct Age method) & 3.02 & 4.01 \\ \hline
 \end{tabular}
\end{table*}

The star formation history that produces a synthetic population with the best fit to the observed 40\,pc sample is drawn from \cite{Mor_2019}, having a $\chi^2_\nu$ 1.2 times lower than the star formation history derived in this work, the next best fit. Because these are single values of $\chi^2_\nu$, and at this stage we are not considering uncertainties on any ingredients in the simulation, these slight differences in $\chi^2_\nu$ are not statistically significant. The systematic uncertainties incorporated in Sect.\,\ref{sec:errors} are required before we can make statistically significant comparisons.


We also ran absolute \textit{G} magnitude distribution simulations with each of the four different assumed star formation history forms and calculated the $\chi^2_\nu$ for each in Table\,\ref{tab:best_chi}. For this method, the best fitting star formation history forms are the constant one of \cite{Cukanovaite_2023}, that of \cite{Mor_2019}, and the star formation history derived from this work which are statistically equivalent. These forms are both a significantly better fit than \cite{Fantin_2019} and our direct age star formation history using this method. The issue of comparing $\chi^2_\nu$ values as with the luminosity function method also applies here; without considering systematic uncertainties, the slight differences between the $\chi^2_\nu$ values are not statistically significant.

These results make it clear that for a fixed set of a \textit{Gaia}-defined stellar sample, astrophysical relations and stellar models, the derived formation history can be different depending on the methodology being used. The most likely reason for this behaviour is that the different methods rely on different subsets of \textit{Gaia} data to reach their conclusion, which are subject to different systematics from both observations and input models. Next we evaluate what those systematics on the stellar formation history are.



\section{Uncertainties}~\label{sec:errors}


So far the synthetic populations have been generated from the fixed ingredients listed in Table\,\ref{tab:ingredients}, the default values. However in order to determine if any star formation history form is a significantly better fit to the 40\,pc population than any other form, or if one of the methods under investigation is better than another, we incorporate uncertainties on these external ingredients. For each of the three methods, the code was ran with each ingredient being allowed to vary around the default values, using a Gaussian distribution unless noted otherwise, as described in the following subsections. 

\subsection{Systematic uncertainties}

\textbf{Initial mass function:} By default, the simulation uses the \cite{Salpeter_1955} initial mass function with $\rho(M) \propto M^{-2.35}$ where $M$ is the initial mass of the star and $\rho(M)$ is the initial mass density function. For the standard deviation, we choose a value of $0.1$ \citep{Weisz_2015, ElBadry_2018, Cunningham_2024} on the exponent to allow the initial mass to vary around the Salpeter value but overlap with different common choices, such as \cite{Kroupa_2001} which uses $\rho(M) \propto M^{-2.3}$. 

A steeper slope to the initial mass function (more negative exponent) means that the simulation will produce more low mass stars. This in turn means the simulated population will have longer main sequence lifetimes on average. A longer main sequence lifetime with the same population age and formation times means shorter white dwarf cooling times; so on average the population will have brighter magnitudes and higher effective temperatures than a population with a shallower initial mass function slope.

\noindent \textbf{Main sequence lifetimes:} We adopt the estimate of a 4.8 per cent uncertainty on main-sequence lifetimes from \citet{Hurley_2000}, independent of mass or metallicity. To incorporate this error for the systematic uncertainties, we include the same shift to all main sequence lifetimes in a given simulation run, drawing the value of the shift from a Gaussian with a standard deviation of 4.8 per cent.

By increasing the main sequence lifetime of a star and assuming the formation history remains the same, the white dwarf cooling time of the star is reduced. This means that on average, longer main sequence lifetimes result in brighter and hotter white dwarfs in the population. 


\noindent \textbf{Population age: } In the original simulation, the age of the population is set to be $10.6$\,Gyr based on \cite{Cukanovaite_2023}, who found a value of $10.6\pm 0.5$\,Gyr for the age of the local 40\,pc population. Fitting the age of the 40\,pc population using the Luminosity Function and the Absolute \textit{Gaia G} Magnitude methods described here returned a wide range of ages, with the 10.6\,Gyr from \cite{Cukanovaite_2023} as very close to the average of our fitting results. Thus, we did not fit the population age in this work, and chose to use 10.6\,Gyr as the default value in all of our simulations. For our evaluation of systematic errors, the population age is drawn from a Gaussian with a mean of 10.6\,Gyr and a standard deviation of 0.7\,Gyr. This standard deviation is reflective of
uncertainties we found when fitting the population age as a free parameter.

By changing the age of the population (and therefore the assumed age of the Galactic disc), the age and the cooling time of the white dwarfs are also affected. An older population age allows for stars that have formed longer ago; this in turn means the white dwarfs have longer cooling ages, are cooler, and have fainter magnitudes at the present day. An older population is on average fainter with cooler effective temperatures. 

\noindent\textbf{Initial-final mass relation:} The initial-final mass relation is derived using the method of \cite{Cunningham_2024}, hence from the 40\,pc white dwarf sample that we also use in this work. 
Since the employed initial-final mass relation is already self-consistent with the \textit{Gaia} temperature, mass and magnitude scales for the 40\,pc sample, and already based on essentially the same astrophysical relations and stellar models as those used in this work, we do not vary this parameter. In other words, the initial-final relation is primarily an internal dependent relation in this work, emerging as a consequence of our particular choices of main-sequence lifetimes, initial mass function and \textit{Gaia} photometric white dwarf masses.

We do not account for individual white dwarf mass scatter around the median initial-final mass relation, due to metallicity, magnetic field or rotation \citep{Cunningham_2024}, although the scatter is expected to be at most a few percent \citep{Hollands_2024}.



\noindent \textbf{Age versus velocity dispersion relation:} As discussed in Sect.~\ref{sec:LF}, older objects have a higher velocity dispersion. Because of this, they are more likely to have left a fixed volume of space close to the Galactic disc plane in the time since they formed. In the direction perpendicular to the disc plane, stars oscillate in their position with time, hence this relation can be characterised by the vertical scale height of the stars of a given age. Our default relation is based on the vertical velocity dispersion of the 40\,pc white dwarfs under study as a function of age \citep{Cukanovaite_2023}. However, both the ages of these white dwarfs, as well as their true velocity dispersion given the lack of radial velocities, remain uncertain. Several different age versus velocity dispersion relations have been published in the literature \citep{Seabroke_2007,Buckner_Froebrich_2014,Cheng_2019}, including using white dwarf samples \citep{Wegg_Phinney_2012,Raddi_2022}. 
To account for this systematic, we apply a multiplicative factor to the gradient of the scale height relation in the age vs velocity dispersion relation, with a mean of 1.0 and a standard deviation of 0.3, sampled from a Gaussian distribution for each simulation run. This provides a wide range of assumed scale heights, reflective of the fact that this parameter is only loosely constrained. The relation always flattens off at 7\,Gyrs due to the lack of kinematic information on stars of this age.

A multiplicative factor below one reduces a star's scale height for any age, which also reduces its probability of having left the sample by the present day. This means that more old, high velocity, faint magnitude white dwarfs remain in the 40\,pc simulated volume, and the population is on average fainter. 

We note that white dwarfs in the local volume also show an age versus velocity dispersion relation in directions within the Galactic disc plane \citep{Raddi_2022}. This suggests that our assumptions that the same number of stars come in and come out of the sample in these directions may be incorrect, and that in fact white dwarfs in the sample may have formed, on average, closer or further away to the Galactic centre \citep{Zubiaur_2024}. However, this question is largely outside of the scope of this work, as we only directly constrain the stellar formation history for objects currently in the local stellar volume -- and not the formation history of the Galactic disc at a specific distance to the Galactic centre -- and therefore we do not need to account for this systematic. \citet{Zubiaur_2024} suggest that 68\,per cent of white dwarfs currently within 100\,pc were formed at less than 1\,kpc from the Sun, although if that interpretation was to change from a different model of radial disc migration, our results could easily be translated to this new model.

\noindent \textbf{Cooling ages:} The white dwarf cooling age models of \cite{Bedard_2020} depend on the envelope composition and mass of the white dwarf. The cooling time in Eqn.\,\ref{eqn:cooling_time} uses these models, from which the effective temperature is subsequently calculated using the same models. The cooling age for the white dwarfs in the direct calculation method is interpolated using mass, envelope composition and effective temperature from the same models. 

White dwarf cooling models calculated by different groups \citep{Camisassa_2016, Camisassa_2019, Bedard_2020, Salaris_2022, Bauer_2023, Pathak_2024} cause ages to vary by several per cent due to uncertainties on core and envelope chemical compositions, atomic opacities and electronic conduction. These uncertainties do not consider the crystallisation and distillation processes which are discussed independently in the next bullet point. Therefore, we consider that the uncertainty on cooling ages is 6 per cent \citep{Cukanovaite_2023,Cunningham_2024, Pathak_2024}.
We apply a multiplicative factor to all cooling times in a given simulation run, drawing the factor from a Gaussian distribution with a mean of 1.0 and a standard deviation of 0.06. Shorter cooling ages predict a brighter and higher temperature white dwarf population on average.

\noindent \textbf{Crystallisation delay:} The standard white dwarf cooling models of \cite{Bedard_2020} do not reproduce the full extent of the cooling delay due to crystallisation and related processes observed in the \textit{Gaia} Hertzsprung-Russell diagram \citep{Tremblay_2019,Blouin_2020,Kilic_2020,Bedard_2024_Qbranch}. Our simulations use an approximation of the additional distillation delay discussed in \cite{Blouin_2021} and assume that any white dwarf that has a crystallised mass fraction (as determined from the cooling models of \citealt{Bedard_2020}) of $\geq0.5$ has experienced an additional cooling delay of 0.5\,Gyr. 
To assess the uncertainty on this component, we employ a random binary choice between including and not including the distillation delay in a given simulation. This uncertainty is similar to the one on the cooling times detailed previously, but is only assigned to older white dwarfs that are undergoing crystallisation.


\noindent \textbf{Binary evolution and stellar mergers:} The simulation uses the evolutionary delays arising from the merger of binary systems from \cite{Temmink_2020} as described in Sect. \ref{sec:LF}. As the delays already include Gaussian scatter, to assess the uncertainties here, the inclusion of mergers is randomly turned off or on for a simulation run. 

In general, 
being a merger product lengthens either the pre-white dwarf lifetime (MS+MS mergers) or white dwarf cooling time (WD+WD mergers), and results in brighter white dwarfs at the present day. Turning off the merger delays therefore causes on average fainter white dwarfs in the population.

\noindent \textbf{Spectral evolution:} 
For the fraction of helium-atmosphere (thin hydrogen layer) white dwarfs, we adopt a standard deviation of 0.1 on the mean of 0.25 and sample from a Gaussian distribution. This accounts for both small number statistics in spectra evolution and additional uncertainties arising from subtypes with magnetic fields and metal pollution that can affect temperature and mass determinations.

The effect of changing the fraction of helium-atmosphere white dwarfs is non-linear: the interaction of composition, mass and cooling age means that He white dwarfs are not consistently brighter than H white dwarfs at the same mass and cooling age, or vice versa. 

\subsection{Individual star uncertainties}

\textbf{Metallicity:} With the knowledge that the solar neighbourhood does not appear to show any age-metallicity relationship \citep{RebassaMansergas_2021} except possibly for the oldest few percent of the sample \citep{Kilic_2019}, we assume that the median metallicity in mass fraction of stars in the simulation is solar regardless of formation time, where $Z_{\odot} = 0.0134$ from \cite{Asplund_2009}. When looking at the uncertainties in this value, metallicity is sampled from a Gaussian with the mean value being solar, $Z>0$ and the standard deviation being 0.0104 \citep{Cunningham_2024}. 


\noindent \textbf{\textit{Gaia} data errors:} When 
simulating the white dwarf mass distribution, we alter the masses of the simulated white dwarfs by adding noise drawn from a normal distribution with a standard deviation of 0.02\,\msolar~in line with the median \textit{Gaia} statistical uncertainty on the masses from \cite{OBrien_2024}. For the luminosity function and absolute $G$ magnitude distribution, we instead propagate the \textit{Gaia} error bars to the observed distributions.

\subsection{Overall uncertainties}

\begin{figure*}
	\includegraphics[width=2\columnwidth]{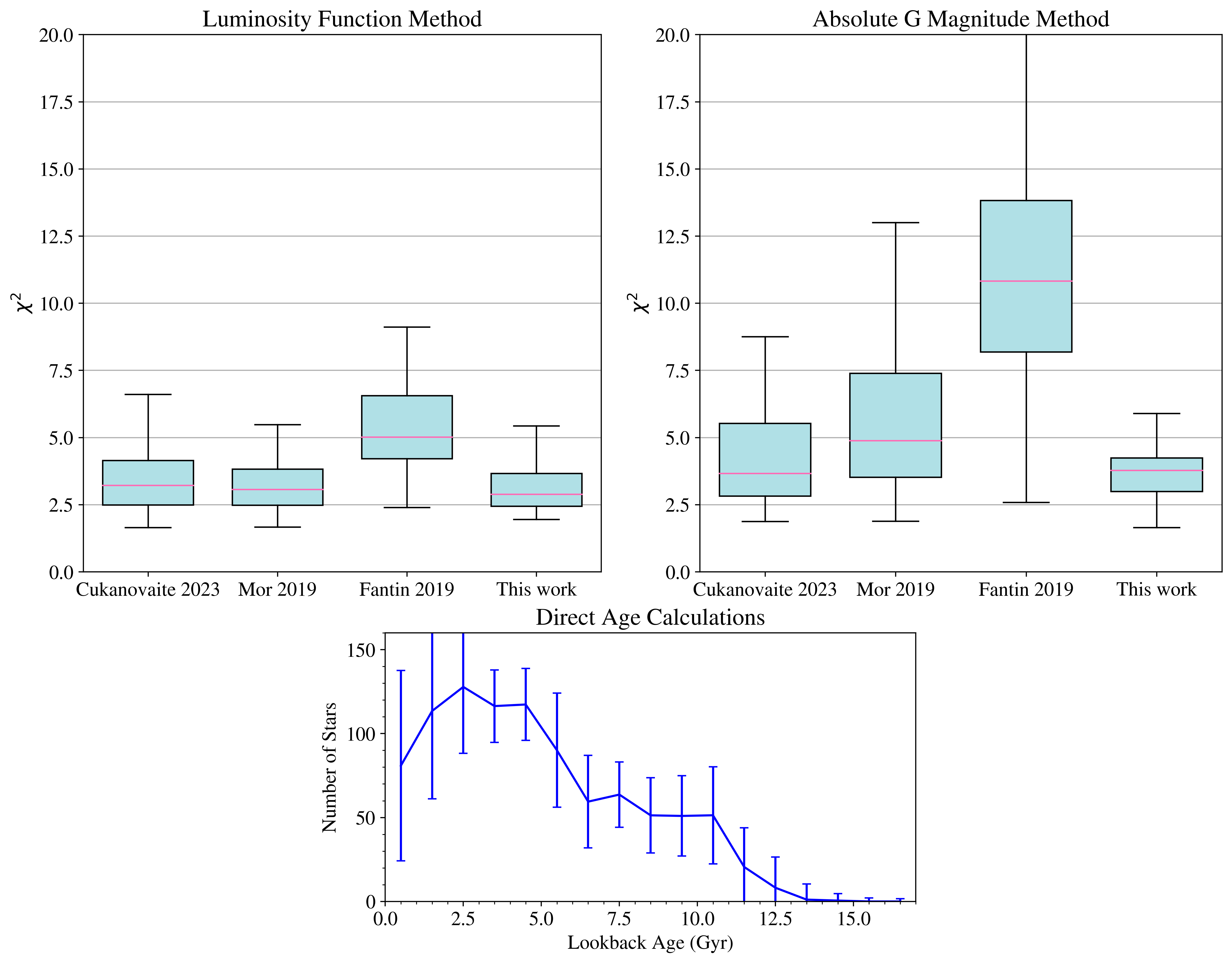}
    \caption{\textit{Luminosity Function method:} Box and whisker plot to show the range of $\chi_\nu^2$ values obtained from 100 runs of the Luminosity Function code including systematic uncertainties. The box extends from the first quartile to the third quartile of the data with a line at the median value (drawn in pink). The whiskers extend from the edge of the box to the furthest data point that lies within 1.5 times the inter-quartile range. Any points past the whiskers are not plotted. 
    \textit{Absolute \textit{G} Magnitude method:} Box and whisker plots to show the range of $\chi_\nu^2$ values obtained from 100 runs of the Absolute \textit{G} Magnitude code including systematic uncertainties. The boxes and whiskers are drawn the same way as before. 
    \textit{Direct age calculations: } The star formation history obtained from the direct age calculation method and the systematic uncertainties on that form. Vertical errors are Poisson errors and the average spread of bin values across the 100 runs added in quadrature when systematic uncertainties and \textit{Gaia} observational errors are applied; Poisson errors on average account for 10 per cent of the uncertainty, systematic uncertainties account for 47 per cent, and \textit{Gaia} observational uncertainties account for 43 per cent. 
    }
    \label{fig:boxplots}
\end{figure*}

We ran a set of 100 simulations using the uncertainties described above, and 100 values of $\chi_\nu^2$ were collected. Fig.\,\ref{fig:boxplots} shows the spread of the 100 $\chi_\nu^2$ values from the Luminosity Function and Absolute \textit{G} Magnitude distribution methods from all four tested star formation histories in a box and whisker plot, as well as the results of the direct age calculations and the uncertainties on those ages when systematic and individual errors are propagated.

By comparing the $\chi_\nu^2$ values we obtain from the default simulation ingredients (shown in Table \ref{tab:best_chi}) with the spread in Fig. \ref{fig:boxplots}, it can be seen that there are no statistically significant differences between any of the four star formation histories: the boxes representing the inter-quartile range (less than $1\sigma$ spread) overlap across the choice of star formation history, with the whiskers representing 1.5 times the inter-quartile range overlapping even further. This shows that no choice of star formation history will give a statistically better result even if the median values of $\chi^2_\nu$ shown in pink do favour certain star formation histories over others.
While it seems that the early peaking star formation history of \cite{Fantin_2019} is not favoured by any of the three methods discussed here, any notable differences in the methods or star formation histories are washed out by the systematic uncertainties. 

The direct age calculations also show large uncertainties due to systematics. In particular, the vertical uncertainty towards the present day makes constraining the height of the recent star formation rate enhancement difficult, and the age uncertainties with older white dwarfs can reach nearly 3\,Gyr, making the onset of star formation difficult to constrain. Despite this, the enhanced star formation rate at later times appears to be significant compared to the reduced rate at earlier times.

By breaking down the uncertainties on the direct age calculations, we determine that Poisson errors account for 3--7\% of the uncertainty, \textit{Gaia} data error accounts for 43--47\% of the uncertainty, and the systematics discussed in this section account for 50\% of the uncertainty. Of that 50\%, cooling age uncertainty is the largest contributor with 16\%, followed by metallicity with 14\%, and binary evolution and stellar mergers with 10\%. The other contributing sources are: age versus velocity dispersion relation, 5\%; initial mass function, 3\%; main sequence lifetimes, 1\%; and crystallisation delay, 1\%.

\section{Reducing Systematic Uncertainties}~\label{sec:reduce_err}

\renewcommand{\arraystretch}{1.5}

\begin{table*}
	\centering
	\caption{Reduced uncertainties from comparison with the white dwarf mass distribution}
	\label{tab:reduced_errs}
	\begin{tabular}{||lll||} 
 \hline
  Ingredient & Original uncertainty & Reduced uncertainty   \\ \hline \hline
  Population age & $\mu = 10.6$\,Gyr, $\sigma =0.7$\,Gyr & $\sigma = 0.5$\,Gyr \\ \hline
  Initial mass function & $\mu = 2.35$, $\sigma = 0.1$ & $\sigma = 0.075$ \\ \hline
  Initial metallicity & $\mu = 0.0134$, $\sigma = 0.0104$ & no change \\ \hline
  Main sequence lifetimes & multiplicative, $\mu = 1$, $\sigma = 0.048$ & $\sigma = 0.035$ \\ \hline
    He-atmosphere WD fraction & $\mu = 0.25$, $\sigma = 0.1$ & $\sigma = 0.065$ \\ \hline
  Merger delays & off/on, equal weighting & weighting off/on changed to 43/57 \\ \hline
  Age vs. kinematic relation & multiplicative to slope of scale height, $\mu = 1$, $\sigma = 0.3$ & no change \\ \hline
  Cooling models & multiplicative, $\mu = 1$, $\sigma = 0.06$ & $\sigma = 0.045$ \\ \hline
  Crystallisation cooling delay & off/on, equal weighting & no change \\ \hline
 \end{tabular}
\end{table*}

The uncertainties on the external ingredients that are used to recover the local stellar formation history are large, multidimensional, and most likely correlated to one another. In other words, for a given \textit{Gaia} white dwarf sample, there is a degeneracy between the luminosity function, absolute $G$ magnitude distribution, functional form of the stellar formation history, and external astrophysical relations. However, it is possible to partially lift that degeneracy by comparing the observed and simulated white dwarf mass distributions. In essence, the degeneracy is partially lifted because the white dwarf mass distribution depends more on the initial mass function and main-sequence lifetimes, but much less on what happens during subsequent white dwarf evolution since white dwarf masses remain constant with time \citep{Tremblay_2016, Cunningham_2024}.  The white dwarf mass distribution is also very well studied and so provides a robust observational sample to simulate.


To simulate the 40\,pc white dwarf mass distribution, we use the same simulations as for the luminosity function and absolute \textit{G} magnitude methods. However, we require fewer steps, models and assumptions, since we assume that all 40\,pc white dwarfs are observed regardless of their effective temperatures \citep{Cunningham_2024}, hence do not calculate a cooling age, effective temperature or distillation delay.

By calculating $\chi_\nu^2$ values between the simulated and observed 40\,pc mass distributions and only collecting the set of input parameters if $\chi_{\nu\mathrm{, mass}}^2 \leq 3$,
we can examine only the simulation runs where we are guaranteed a good match to the white dwarf mass distribution. This is because $\chi_{\nu}^2 = 3$ is approximately three times larger than the minimum value we find. It is worth noting that the specific choice of 3 does not affect the outcomes here: limiting $\chi_{\nu}^2$ to any value 2--6 achieves a very similar limit on the uncertainty parameter space. We then take a closer look at the values for the ingredients that passed such quality criteria.

For a constant star formation history, we collected 100 simulation runs that also meet our $\chi_{\nu\mathrm{, mass}}^2 \leq 3$ criteria. Table\,\ref{tab:reduced_errs} compares the input sampled uncertainties with the spread of output values for each ingredient, i.e. our \textit{reduced} uncertainties.

The results shown indicate that the original uncertainties on population age, initial mass function, main sequence lifetimes, helium-and atmosphere fraction, the inclusion of merger delays, and cooling ages were overestimated compared to what is allowed by the 40\,pc white dwarf mass distribution. The ingredients that experienced no change are typically those that do not affect the shape of the white dwarf mass distribution. We also note that the initial-final mass relation is fixed in our study and based on most of the same ingredients as those used in this work. While this biases the results towards the selected median slope of the initial mass function, model main-sequence lifetimes and population age, we make no claim of constraining the absolute value of these three quantities in this work, and instead the self-consistent initial-final mass relation allows us to constrain the size of the error bars on all ingredients.

\begin{figure*}
	\includegraphics[width=2\columnwidth]{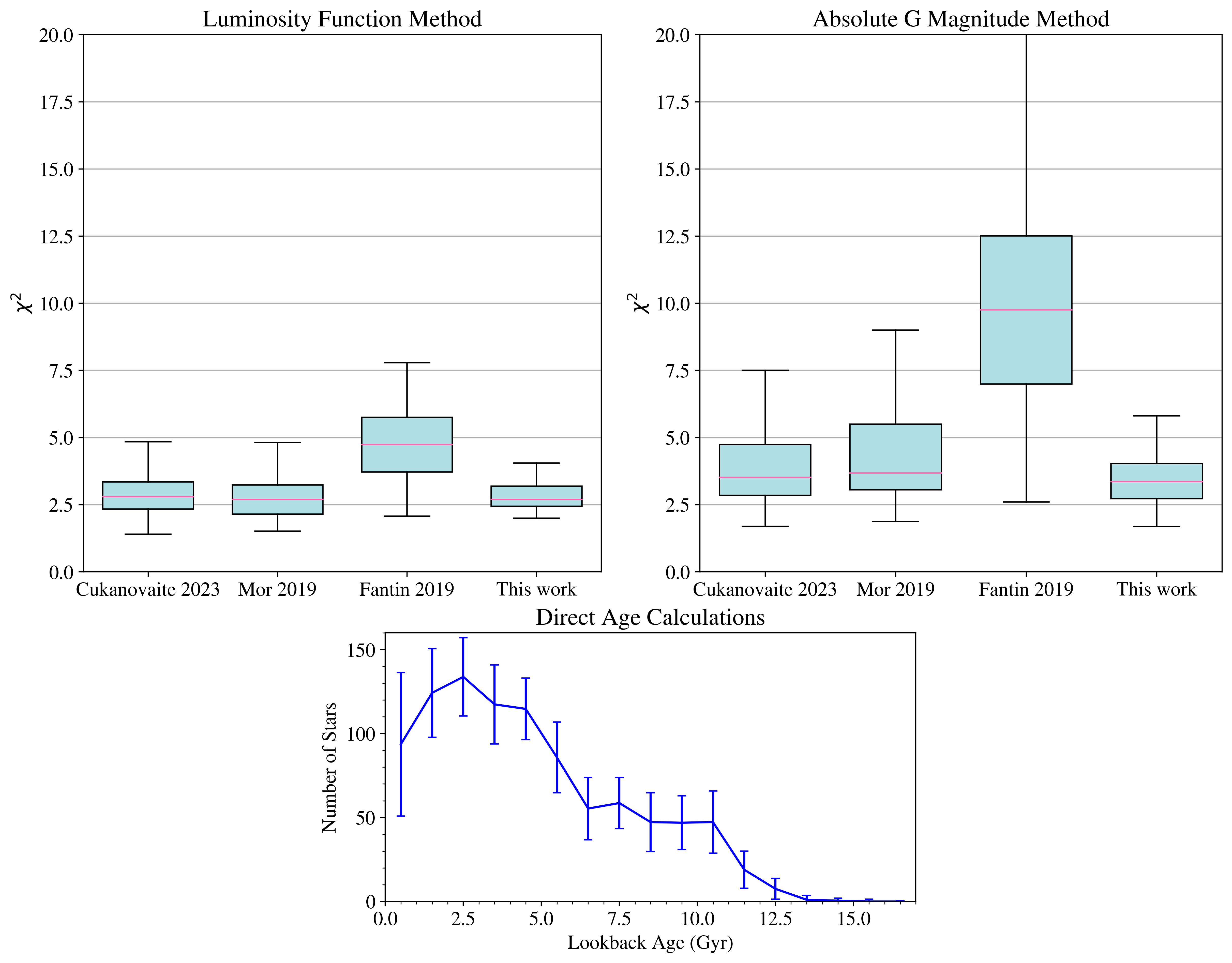}
    \caption{Same as Fig.\,\ref{fig:boxplots}, but with the reduced systematic uncertainties given in Table\,\ref{tab:reduced_errs} after the additional step of comparing the population synthesis with the observed 40\,pc mass distribution.
    }
    \label{fig:boxplots_reduced}
\end{figure*}

We ran the code for the luminosity function and absolute \textit{G} magnitude methods and different formation histories again as in Fig.\,\ref{fig:boxplots} with the reduced uncertainties as determined by restricting the $\chi^2_{\nu, \mathrm{mass}}$. Fig.\,\ref{fig:boxplots_reduced} shows that this does reduce the spread and location of the box and whisker plot as expected. 


\subsection{Sub-samples of massive white dwarfs}\label{sect:high_mass}

Another way to test systematic uncertainties is to look at the stellar formation history for white dwarfs above a certain mass threshold, with the higher mass white dwarfs having smaller main sequence lifetimes, and smaller uncertainties on their main sequence lifetimes owing to the steep inverse correlation of this parameter with mass. In the past, studies have used a cut-off at either 0.63\,\msolar\ \citep{Heintz_2022} or 0.9\,\msolar\ \citep{Isern_2019}. We note, however, that recent studies have also shown that higher mass white dwarfs may have higher uncertainties on their cooling ages from a higher fraction of stellar mergers \citep{Temmink_2020,Bedard_2024_Qbranch, Jewett_2024}.

\cite{Cheng_2019} showed that 5--9\,per cent of ultra-massive white dwarfs (1.08--1.23\,\msolar) experience a cooling delay of at least 8\,Gyr, which produces the high mass portion of the Q branch in the Hertzsprung-Russell diagram. \cite{Bedard_2024_Qbranch} demonstrated that this long delay is the result of distillation in white dwarfs with high core $^{22}$Neon abundances, arising from either high-metallicity progenitors or stellar mergers. However, distillation may also occur in white dwarfs descending from solar-metallicity progenitors and thus having a standard $^{22}$Neon content, leading to much shorter delays \citep{Blouin_2021, Venner_2023}. Our simulations consider the latter scenario by assuming that all crystallising white dwarfs experience an additional 0.5\,Gyr distillation delay to their cooling. However, the small population of extra-delayed massive white dwarfs cannot be probed using the 40\,pc sample, as the $>$0.9\,\msolar\ sub-sample only contains 66 white dwarfs. Hence, small number statistics prevents us from testing different delay scenarios and resulting formation histories for that mass range. 


We can still test the impact of different distillation delays on the full 40\,pc population. We find that extrapolating the scenario of \cite{Bedard_2024_Qbranch} to all white dwarf masses (that is, 7\,per cent of all white dwarfs experience a cooling delay of 8\,Gyr) makes no significant differences to the predicted luminosity function and absolute $G$ magnitude distribution. This is largely a consequence of crystallising white dwarfs of all masses having a wide range of absolute magnitudes. The most pronounced effect of the distillation delay is seen instead in the predicted two-dimensional mass versus $T_{\rm eff}$ distribution \citep{Kilic_2025_100pc}. However, we also find in that case that the two choices of distillation delays discussed in this section -- 0.5\,Gyr for all or 8\,Gyr for 7 per cent of white dwarfs -- are largely degenerate. The main difference between the two delay types concerns the inferred population age, as a systematic 0.5\,Gyr delay results in fewer faint simulated white dwarfs compared to the case of a longer delay for a fraction of white dwarfs. However, the uncertainties on the inferred age of the population are of the same order as the distillation delay, making any conclusion difficult.

Another way to test distillation delay effects is using kinematics. However, the large majority of white dwarfs in the 40\,pc sample have disc kinematics appropriate for their estimated ages \citep{McCleery_2020, Kilic_2020, Cukanovaite_2023}. Since the average total age is large, at $\approx$ 5\,Gyr, a moderate cooling delay of several Gyrs will only increase the total age by a modest factor, hence such delay is not expected to have a strong signature on the kinematics of the full 40\,pc sample. As a consequence, observed kinematics offer weak constraints on the distillation delay of average mass white dwarfs \citep{Kilic_2020}. This is different to the massive sub-sample where the distilling white dwarfs have short main-sequence lifetimes and cooling ages, hence a proportionally much larger distillation delay. 

To minimise uncertainties from stars having long and uncertain main-sequence lifetimes, we investigated the effect of applying a cut-off of $>$0.63\,\msolar\ in white dwarf masses in both the population synthesis and observed sample, at the expense of having $\approx$40\% larger Poisson errors. This experiment reveals that the constant star formation history form remains the best fitting star formation history on the smaller sub-samples, the same as for the full 40\,pc sample. This smaller sample does produce higher $\chi^2_\nu$ than the full 40\,pc sample, and the systematic uncertainties remain large, making it difficult to assess whether one star formation history is significantly better than the others in much the same way as with the full simulations. We conclude that high mass sub-samples are not key to extracting a more precise stellar formation history.

\section{Conclusions}~\label{sec:disc}

We have investigated three methods of determining the star formation history from a white dwarf population: the luminosity function, absolute $G$ magnitude distribution and direct age calculations. 
We find that the systematic uncertainties on main-sequence, post main-sequence and white dwarf models, as well as astrophysical relations such as the initial mass function and age versus velocity dispersion relation, dominate over any underlying differences between the methods. All three methods and their best fit star formation histories agree with one another within errors. This suggests that the choice of which method to use should be influenced by which data are available to use. For instance, the absolute $G$ magnitude method only requires a magnitude and a parallax for each white dwarf, and does not require individual white dwarf temperatures or masses. The method may therefore be better suited for larger white dwarf samples that do not have extensive follow-up spectroscopy. As the systematic uncertainties on any of the three methods are large, the focus of future work should be on reducing uncertainties on input stellar and Galactic models where possible to better constrain simulations of white dwarf populations. In particular, we suggest a focus on reducing the uncertainties on white dwarf cooling times, merger delays, the local stellar metallicity distribution, and the scale height of the Galactic disc as a function of age.


We explored four forms of star formation histories of the Galactic disc for the last 10.6\,Gyr throughout this paper; a first one derived from our direct age calculations, \citet{Fantin_2019, Mor_2019,Cukanovaite_2023}. These are 1) peaking at recent times, 2) peaking at old times, 3) double peaked or 4) constant, respectively, with the maximum and minimum formation rates at any time different at most by a factor of three. 

We note that the star formation history of \cite{Fantin_2019} which peaks at old times does give a worse fit with the 40\,pc sample using both the Absolute \textit{G} Magnitude and Luminosity Function methods, and disagrees with the Direct Age method which finds a star formation history that peaks at recent times. This worse fit is not statistically significant, however, due to the size of the systematic uncertainties.

The star formation history derived in this work via the Direct Age method agrees with the constant star formation history of \cite{Cukanovaite_2023} within $2\sigma$, and both star formation histories have similar $\chi^2_{\nu}$ fits with the 40\,pc sample, making it difficult to quantitatively say which is favoured by the 40\,pc sample.


Due to this lack of a statistically significant best fit across the three fitting methods and four forms of stellar formation histories explored, we agree with the conclusion of \cite{Cukanovaite_2023} that a constant star formation rate is advised as the most simple, default scenario for simulations of the 40\,pc sample and more broadly for the star formation rate of the local Milky Way. We also present tentative evidence that star formation histories with early peaks should be avoided. These results will hopefully guide future simulations and interpretations of volume-limited stellar samples in the Solar neighbourhood.



\section*{Acknowledgements}

This research received funding from the European
Research Council under the European Union’s Horizon 2020 research and
innovation programme number 101002408 (MOS100PC). AB is a Postdoctoral Fellow of the Natural Sciences and Engineering Research Council (NSERC) of Canada. TC was supported by NASA through the NASA Hubble Fellowship grant HST-HF2-51527.001-A awarded by the Space Telescope Science Institute, which is operated by the Association of Universities for Research in Astronomy, Inc., for NASA, under contract NAS5-26555. This work has made use of data from the European
Space Agency (ESA) mission Gaia (\url{https://www.cosmos.esa.int/gaia}), processed by the Gaia Data Processing and Analysis
Consortium (DPAC, \url{https://www.cosmos.esa.int/web/gaia/dpac/consortium)}. Funding for the DPAC has been provided by
national institutions, in particular the institutions participating in the
Gaia Multilateral Agreement.

\section*{Data Availability Statement}

The observational data used in this article are published in \cite{McCleery_2020,GentileFusillo_2021} and \cite{OBrien_2024}. The results derived in this article will be shared on a reasonable request to the corresponding
author.




\bibliographystyle{mnras}
\bibliography{aamnem99,bib_emily,bpass,bedard} 






\bsp	
\label{lastpage}
\end{document}